\documentclass[12pt]{spieman}  
\usepackage{amsmath,amsfonts,amssymb}
\usepackage{graphicx}
\usepackage{setspace}
\usepackage{tocloft}
\usepackage{lineno}
\usepackage{bm}
\usepackage{textcomp}
\usepackage{xcolor}
\usepackage{mathtools }
\linenumbers
\title{  Bernoulli generalized likelihood ratio test for signal detection from photon counting images}

\author[a,*]{Mengya~(Mia)~Hu}
\author[b]{ He~Sun}
\author[a]{Anthony~Harness}
\author[c]{N.~Jeremy~Kasdin}
\affil[a]{Department of Mechanical and Aerospace Engineering, Princeton University, Princeton, NJ, 08544, USA}
\affil[b]{Department of Computing and Mathematical Sciences,
 California Institute of Technology, Pasadena, CA, 91125, USA}
\affil[c]{College of Arts and Sciences, University of San Francisco, San Francisco, CA, 94117, USA}

\cftpagenumbersoff{figure}
\cftpagenumbersoff{table} 
\begin{document} 
\nolinenumbers
\maketitle
\begin{abstract}
Because exoplanets are extremely dim, an Electron Multiplying Charged Coupled Device (EMCCD) operating in photon counting (PC) mode is necessary to reduce the detector noise level and enable their detection. Typically, PC images are added together as a co-added image before processing. We present here a signal detection and estimation technique that works directly with individual PC images. The method is based on the generalized likelihood ratio test (GLRT) and uses a Bernoulli distribution between PC images.   The Bernoulli distribution is derived from a stochastic model for the  detector, which accurately represents its noise characteristics. We show that our technique outperforms a previously used GLRT method that relies on co-added images under a Gaussian noise assumption and two detection algorithms based on signal-to-noise ratio (SNR). Furthermore, our method provides the maximum likelihood estimate of exoplanet intensity and background intensity while doing detection.  It can be applied online, so it is possible to stop observations once a specified threshold is reached, providing confidence for the existence (or absence) of planets.  As a result,  the observation time is efficiently used. Besides the observation time, the analysis of detection performance introduced in the paper also gives quantitative guidance on the choice of imaging parameters, such as the threshold. Lastly, though this work focuses on the example of detecting point source, the framework is widely applicable. 
\end{abstract}


{\noindent \footnotesize\textbf{*}Mengya~(Mia)~Hu,  \linkable{ mengyah@princeton.edu} }

\begin{spacing}{2}   

\section{Introduction}
\label{sec:intro} 
\begin{figure}[ht!]
\center
\includegraphics[width=0.99 \textwidth,trim= 0 12cm  0 0 ,clip]{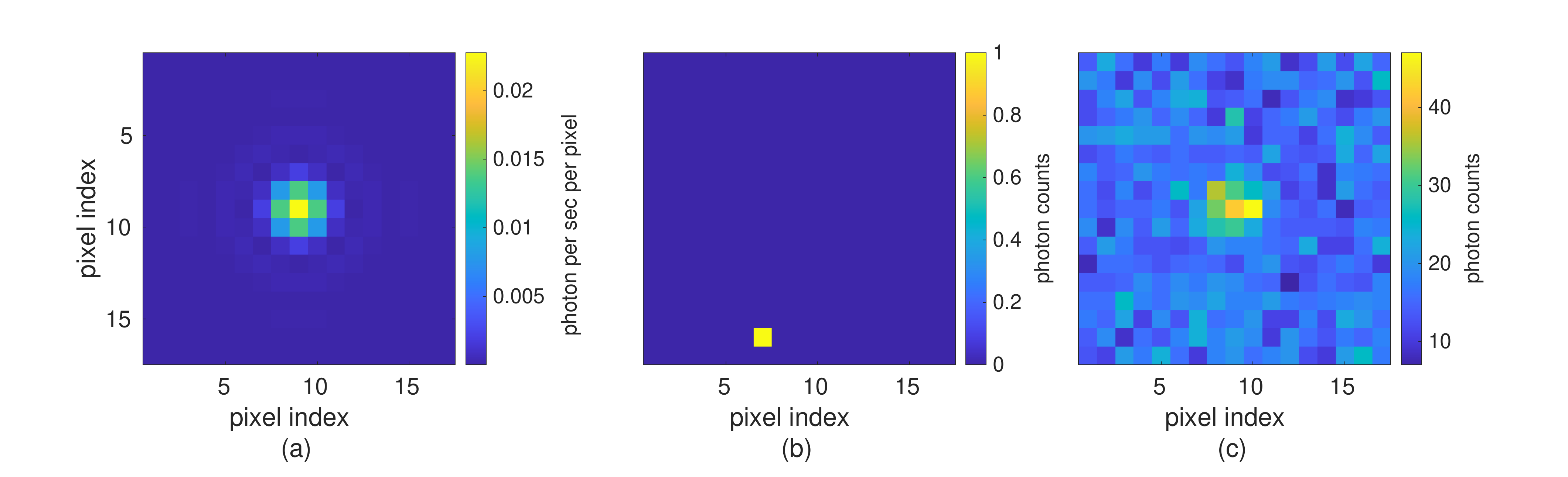}
\caption{(a) Simulated point spread function (PSF) for The Nancy Grace Roman Space Telescope without a starshade for a 1e-8 Jansky(Jy) light source, 0.021 arcsec/pixel. The parameters and simulation process are described later in Secs.~\ref{sec:EMCCD}~and~  \ref{sec:whole}. (b) A photon counting (PC) image of (a) with 1s integration time. (c) A co-added image
from 2000 sequential PC images.}
\label{fig:Image}
\end{figure}

Direct imaging of exoplanets is challenging; the flux ratio between an Earth-like exoplanet and its sun-like host star is around $10^{-10}$ in reflected light at visible wavelengths\cite{Traub2010}. A starshade or internal coronagraph can suppress the starlight and leave only the planet's light to be detected; however, the planets are extremely faint and detecting them is still a challenge. An Earth-like planet ranges from 28$^\text{th}$ to 30$^\text{th}$ magnitude or fainter. As a result, the signal can be smaller than the camera read noise. An Electron Multiplying Charged Coupled Device (EMCCD) can alleviate this problem by amplifying the signal in an electron-multiplication (EM) register, thus reducing the effective readout noise to less than one electron\cite{PCmodel}. Unfortunately, at the same time, a new noise is introduced --- the multiplicative noise from the amplification process. This can be overcome by operating in photon counting (PC) mode. PC Mode reports a value of 1 or 0 in each pixel for each integration time by thresholding the value at the final stage.  The value reported in the pixel is one if the number of electrons in a pixel is bigger than a chosen threshold and zero otherwise. The binary value only indicates the existence of photons in the pixels during the integration time but does not reveal the exact number of those photons, so we need to choose the exposure time such that the expected photon count in any pixel is much less than one\cite{PCtime}. Examples of simulated PC images are shown in Fig.~\ref{fig:Image} (all the simulations mentioned in this work use a 1 second integration time) and details on the generation of simulated images can be found in Ref.~\citenum{Mia1}. 

Operation of an EMCCD in photon counting mode is fairly new and thus  techniques are still developing. Available literature on the design and characteristics of EMCCD detectors can be found in Refs.~\citenum{PC1,PC2,PC3,PC4,PC5,detector1,detector2,PCmodel}. Previous work on image processing focuses mainly on image stacking and Bayesian estimation methods\cite{PCs1,PCs2}, which are applied to co-added PC images rather than designed for individual PC images\cite{vip,KLIP}. As those methods generally require a high signal-to-noise-ratio, large numbers of PC images are needed, which takes a long observation time. In our previous work, we presented an alternative methodology for co-added PC images\cite{Mia} which can efficiently detect even weak signals automatically. However, all these methods do not provide theoretical guidance on how to choose the integration times and the total number of PC images to combine into one co-added image.

In this work, we utilize a statistical model for the EMCCD to obtain a relationship between the detection probability and the photon rate. We use this distribution to formulate a Bernoulli distribution for the values of the same pixel on different PC images. Then, detection and estimation is performed.  Consequently, detection and accurate intensity estimation can be achieved at low signal levels. We begin this paper with a description of the detector model, followed by the methodology of the   Bernoulli Generalized Likelihood Ratio Test (BGLRT) , and an application example. We finish with a comparison of the performance of our previous method, GLRT using co-added images\cite{Mia}, and the new   BGLRT method using PC images directly. This work builds on the previous work introduced in Ref.~\citenum{seqGLRT}  , where the method is called the Sequential Generalized Likelihood Ratio Test. We change the method name to Bernoulli Generalized Likelihood Ratio Test to better reflect that the improvement of performance is from the usage of an accurate model based on the Bernoulli distribution for the detector. 

\section{A stochastic model for EMCCDs in photon counting mode}
\label{sec:EMCCD} 
Hirsch et al.\cite{PCmodel} present a detailed imaging process for an EMCCD and build a statistical model for each stage of the imaging process. Their statistical model is built as follows\cite{PCmodel}. First, incident photons follow a Poisson process. Second, the EM register is represented by a gamma distribution. Third, readout noise is added using a Gaussian distribution. Fourth, a threshold is applied to give a binary result.

Hirsch et al. use some approximations to simplify  and arrive at the equation for the probability distribution for the number of electrons in a pixel, 
\begin{equation}
        \begin{aligned}[b]
        p(n_{ic})&=
        \begin{cases}
            \frac{1}{\sqrt{2\pi} \sigma} exp(-\lambda-\frac{  ( n_{ic}   - B_{PC})^2}{2\sigma^2})+\frac{2}{g}F_\chi(2\lambda;4,\frac{2 n_{ic}}{g}), &  n_{ic}>0 \\
            \frac{1}{\sqrt{2\pi} \sigma} exp(-\lambda + \frac{   B_{PC}^2}{2\sigma^2}), &   n_{ic}=0,
        \end{cases}
\end{aligned}
\label{equ:prob1}
\end{equation}
 where $\lambda$ is the expected number of input electrons for the EM register,
   
 \begin{equation}
 \lambda = s\times q \times t + d \times t + CIC;
         \label{equ:lam}
\end{equation}

$s$ is the photon rate; $q$ is the quantum efficiency; $t$ is the integration time; $d$ is the dark current; $CIC$ is the clock induced charge; $n_{ic}$ is the number of counts after the EM register, including readout noise; $p(n_{ic})$ is the probability of counts $n_{ic}$; $\sigma$ is the standard deviation of the readout noise; $g$ is the EM gain; $B_{PC}$ is the bias in PC mode; $F_\chi(2\lambda;4,\frac{2 n_{ic}}{g})$ is the non-central $\chi^2$ distribution for $2\lambda$ with 4 degrees of freedom and the non-centrality parameter $\frac{2 n_{ic}}{g}$. Units for these parameters are shown in Table~\ref{table:1}.
 
  An EMCCD decreases the read-out noise to sub-electron level by amplifying the signal before readout. However, the amplification process introduces an excess noise factor (ENF) that reaches $2^{\frac{1}{2}}$ at high gains\cite{ENF}. The effect on the SNR is the same as if the quantum efficiency of the EMCCD were halved\cite{Daigle09}. This can be overcome completely by operating in photon counting (PC) mode. PC mode applies a single threshold at the final stage and reports a binary result. The pixel registers a detected photon if its count level is higher than the threshold, and reports no photon otherwise. This solution alleviates the ENF without making any assumption on the signal's stability across multiple images\cite{Daigle09}.

\begin{figure}[ht!]
\centering
\includegraphics[width=0.99 \textwidth,trim= 0 10cm 0 12cm ,clip]{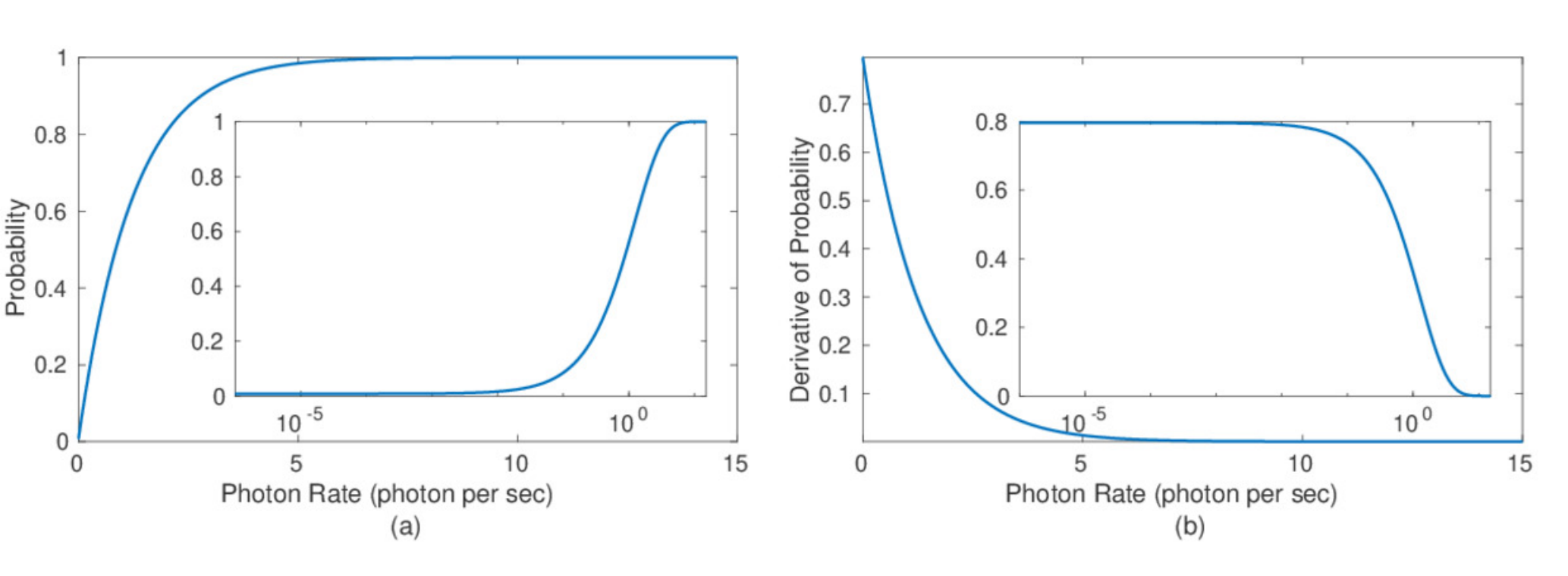}
\caption{Detection probability and its derivative. (a) Detection probability as a function of photon flux, $f(s)$, calculated by Eq.~(\ref{equ:prob2}), where $s$ is in units of photons per second. (b) Derivative of the detection probability. The inserts are the same plots shown with a logarithmic abscissa.  We use the detector parameters values shown in Table~\ref{table:1}, which are similar to the parameter values for the Roman Telescope\cite{detector} and assume a starshade for starlight suppression.}
\label{fig:countingProb}
\end{figure}

Combining models for all the stages described above produces the final model, which calculates the probability of getting a value of 1 in a detector pixel given a certain incident intensity.  Thus, with a threshold of $B_{PC} + T\times \sigma$, set by the threshold parameter $T$, the probability of detecting a  value of one is:
\begin{equation}
f(s) =\int_{B_{PC} + T\times \sigma}^{\infty} p(n_{ic})\, d n_{ic}.
    \label{equ:prob2}
\end{equation}
  where $f(s)$ is shortened for $f(s;q,t,CIC,d,g,B_{PC},\sigma,T)$ (the detector parameters are not shown explicitly). The value in each pixel in each image only has two outcomes: either 1 or 0. The probability of getting  a value of 1, $f(s)$, is decided by the ground-truth of the flux, $s$. As the flux in each pixel is fixed for stable observations, the probability is fixed. That is to say, the measurement in each pixel in each image satisfies a Bernoulli distribution (the Bernoulli distribution is simply the probabilities of “heads” (p) and “tails” (1-p) in a weighted coin flip). In this paper, we assume  that all the detector parameters such as integration time are fixed and the values are shown in Table~\ref{table:1}, so for each pixel, the imaging result follows a  Bernoulli distribution whose probability of value 1 is only related to the flux value  in the pixel. Fig.~\ref{fig:countingProb} shows results for the detection probability given different flux levels and the calculated derivative of this probability. We use the detector parameters values shown in Table~\ref{table:1}, which are similar to the parameter values for the Nancy Grace Roman Space Telescope\cite{detector} and assume a starshade for starlight suppression. The detection probability increases as the flux increases, where a very small flux or a very large flux tend to give deterministic zero or one measurements. 

 The choice of imaging parameters (integration time, PC threshold, etc.) helps efficiently detect signals. The binary value only indicates the existence of photons in a pixel during the integration time but does not reveal the exact number of the photons. To utilize the photons efficiently, the exposure time is chosen so that the expected photon count in any pixel is less than 1.  To decide a threshold for PC, we need to also pay attention to signal intensity and integration time, because they have similar influence on the result. Suppose that we have decided on the integration time and the expected range of signal intensity. To determine the imaging parameters, we can directly utilize the area under the receiver operating characteristic (ROC) curve (AUC) (Details of ROC and AUC are in Sec.~\ref{sec:comp}.), which is a number reflecting the overall detection performance. For one threshold, we can calculate the average of the AUCs for different signal intensities in the expected intensity range. The threshold with the largest average AUC should be chosen.

\begin{table}[h!]
\centering
\caption{EMCCD detector parameters}
\begin{tabular}{ |p{6.5cm}|p{1.3cm}|p{2cm}|p{3.5cm}|  }
 \hline
 Parameter & Symbol & Value & Unit\\
 \hline
 Quantum efficiency & $q$ & 1 & $ph/e^-$\\
 Integration time & $t$ & 1 & $s$\\
 Clock-induced charge   & $CIC$    &  0.01 & $e^-pixel^{-1}frame^{-1}$\\
 Dark current & $d$ & $2\times10^{-4}$ & $e^-pixel^{-1}s^{-1}$\\
 Electron-multiplying gain & $g$ & 2500 & $-$ \\
 PC Bias & $B_{PC}$ & 200 & $e^-pixel^{-1}frame^{-1}$\\
 Standard deviation of readout noise & $\sigma$ & 100 & $e^-pixel^{-1}frame^{-1}$\\
 Threshold Parameter & $T$ & 5.5 & $-$ \\
 \hline
\end{tabular}
\label{table:1}
\end{table}

\section{  Bernoulli generalized likelihood ratio test}

The Generalized Likelihood Ratio Test (GLRT) is a powerful tool for detection.  The detection problem can be seen as deciding between the null hypothesis, $H_0$, that there is no signal, and the alternative hypothesis, $H_1$, that there is a signal. GLRT uses the ratio of the likelihood of observing the data under $H_1$ to the likelihood of observing the data under $H_0$ as the test statistic. If the ratio is large enough, we will decide $H_1$. Compared to other likelihood ratio tests, the GLRT can include hypothesis tests which have unknown parameters in the two hypothesis. GLRT use maximum likelihood estimation (MLE) to estimate the unknowns and then calculates the maximum likelihood ratio.      In other words, the MLEs of the unknown parameters such as  signal intensity and background intensity, under both positive and negative hypotheses, are first required to be calculated. Then, a likelihood ratio test using the MLEs can be calculated and applied to decide whether a signal has been detected. Our previous work approximated the noise in  co-added images by a Gaussian distribution\cite{Mia}. In this work, we directly use the detector's probability function derived in the previous section,  which does not rely on co-added images with a large number of PC images. As a result, measurements in each pixel in each image follow a Bernoulli distribution. 

\subsection{Signal estimation} \label{sec:estimation}
  In this subsection and Sec.~\ref{sec:detection}, we examine an image area that has the size of PSF's  central core (the main part of a signal). We will explain how to apply the method on the whole image, which is normally larger than the PSF core, in Sec.~\ref{sec:whole}.  We begin with the simple model of the signal received by the detector, 
\begin{equation}
\label{eq:signal}
 \bm{s} = \alpha  \bm{x} + \beta  \textbf{1},
\end{equation}
where $\bm{x}$ is a reference point spread function (PSF) of the telescope; $\alpha$ is the intensity of the planet relative to the   source intensity of the reference PSF; $\beta$ is the background light (such as dust and light leakage from starshade defects),  and $\textbf{1}$ is a  column vector where each element equals one. We stack pixel values in the target area into a column vector for easier mathematical manipulation. That is to say,  $\bm{s}$ and $\bm{x} $ are column vectors. 

\begin{figure}[ht!]
\centering
\includegraphics[width=0.53 \textwidth,trim= 0 0cm 0 0cm ,clip]{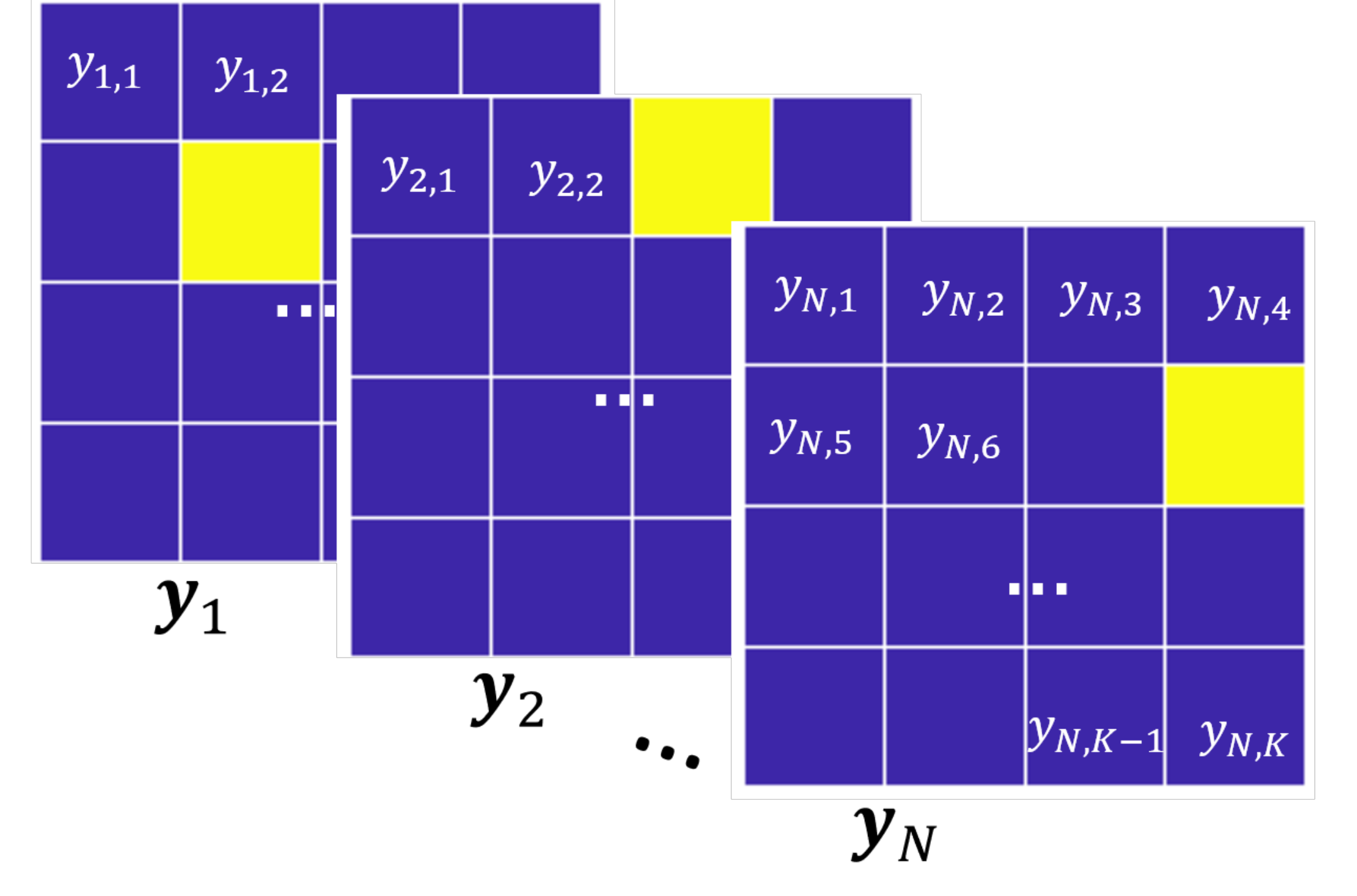}
\caption{  Illustration for the variables in the model. $K$ is the number of pixels in each PC image (with the size of PSF core). $N$ is the number of PC images. }
\label{fig:KN}
\end{figure}

  In real life, due to the randomness of photons and the detector noise, we wouldn`t be able to observe $\bm{s}$ directly, but rather collect the noisy image output from the detector. Assume that $N$ photon counting images have been collected, which is denoted as $\bm{y} = \{\bm{y}_1, \cdots, \bm{y}_N\}$. Given the  PSF shape, $ \bm{x} = (x_1, \cdots, x_K)^T$, where the subscripts of $x$ represent the pixel indices and $K$ is the total number of   pixels in a PSF,    the conditional probability of an image $\bm{y}_n = (y_{n,1},...,y_{n,K})^T$ is
\begin{equation} \label{eq:probI}
  p(\bm{y}_n | \alpha, \beta) = \prod_{k=1}^K  [1 - f(\alpha x_k + \beta)]^{1-y_{n, k}} f(\alpha x_k + \beta)^{y_{n, k}} ,
\end{equation} 
  where the function $f(\cdot)$ is the probability function of detecting a one value in Eq.~(\ref{equ:prob2}). The measurement in each pixel satisfies a Bernoulli distribution, and $[1 - f(\alpha x_k + \beta)]^{1-y_{n, k}} f(\alpha x_k + \beta)^{y_{n, k}}$ is the probability for the measurement in the $k$-th pixel ( for PC images, $y_{n, k}$ is either one or zero). The probability for the whole image is simply the product of the probability of every measurement. The interesting part is that the probabilities in different pixels are correlated because of the underlying signal model in Eq.~\ref{eq:signal}. Fig.~\ref{fig:KN} illustrates the meaning of the indices in the equation. 

  For this work, as a demonstration of the detection method, we use a PSF template as the underlying signal model. However, different signal models can be used and the later discussion of the detection method is still valid; the only difference would be detecting signal with different templates. This provides the   Bernoulli GLRT with a broad applicability to various observation scenarios.

   Furthermore, the conditional probability of $N$ images, $\bm{y} = \{\bm{y}_1, \cdots, \bm{y}_N\}$, can be written as
\begin{equation} \label{eq:prob}
\begin{aligned}
   L(\bm{\theta}) &    = p(\bm{y} | \alpha, \beta) \\
&   = \prod_{n=1}^N p(\bm{y}_n | \alpha, \beta) \\
&   = \prod_{n=1}^N \prod_{k=1}^K  [1 - f(\alpha x_k + \beta)]^{1-y_{n, k}} f(\alpha x_k + \beta)^{y_{n, k}} \\
&   = \prod_{k=1}^K [1 - f(\alpha x_k + \beta)]^{N- \sum_{n=1}^N  y_{n, k}} f(\alpha x_k + \beta)^{\sum_{n=1}^N  y_{n, k}}\\
&   = \prod_{k=1}^K [1 - f(\alpha x_k + \beta)]^{N_{0, k}} f(\alpha x_k + \beta)^{N_{1, k}} ,
\end{aligned}
\end{equation} 
where   $\bm{\theta}$ is the simpler vector representation for the two unknown parameters $\binom{\alpha}{\beta} $ , $N_{0, k}=N- \sum_{n=1}^N  y_{n, k} $ and $N_{1, k}=\sum_{n=1}^N  y_{n, k}$ respectively represent the frequency of zero and one occurring in $N$ measurements of the $k$-th pixel. The equality $N= N_{0,k}+N_{1,k}$ holds for every pixel.   As the data $\bm{y}$ is known and the parameters $\bm{\theta}$ are unknown, this probability function is a likelihood function for the unknown parameters (so it is denoted as $L(\bm{\theta})$ above ).

Taking the natural logarithm of both sides of Eq.~(\ref{eq:prob}), the log-likelihood of the series of measurements is
\begin{equation} \label{eq:likelihood}
l(\bm{\theta})   = ln(L(\bm{\theta}))  = \sum_{k=1}^K N_{0, k} ln [1 - f(\alpha x_k + \beta)] + N_{1, k} ln [f(\alpha x_k + \beta)] .
\end{equation}
To estimate $\alpha$ and $\beta$, we can conduct a MLE based on Eq.~(\ref{eq:likelihood}) by taking the derivatives with respect to $\alpha$ and $\beta$ and setting them equal to zero,
\begin{equation} \label{eq:gd}
\begin{aligned}
\frac{\partial l(\alpha, \beta)}{\partial \alpha} &= N \sum_{k=1}^K x_k f^{\prime}(\alpha x_k + \beta) \frac{N_{1, k} - N f(\alpha x_k + \beta)}{N f(\alpha x_k + \beta) [1-f(\alpha x_k + \beta)]} = 0,\\
\frac{\partial l(\alpha, \beta)}{\partial \beta} &= N \sum_{k=1}^K f^{\prime}(\alpha x_k + \beta) \frac{N_{1, k} - N f(\alpha x_k + \beta)}{N f(\alpha x_k + \beta) [1-f(\alpha x_k + \beta)]} = 0,
\end{aligned}
\end{equation}
  where $f^{\prime}(s) =\frac{df(s)}{ds} $. These equations can be solved for estimates of $\alpha$ and $\beta$ using a gradient descent method.  We denote the solution, i.e., the MLE, with a hat, such as  $\hat{\alpha}$ and $\hat{\beta}$. As can be seen, the derivatives are weighted summations of the difference between the measurements and the distribution means, where $N f(\cdot)$ and $N f(\cdot) [1-f(\cdot)]$ are, respectively, the mean and variance (each measurement satisfies a Bernoulli distribution). According to Fig.~\ref{fig:countingProb} (b), the function  $f^{\prime}(s)$ is zero when $s \rightarrow +\infty$. That means that the method neglects the difference between different intensities that are too large in the estimation since the photon counting detector always gives 1 in these cases.

  In the problem of parameter estimation, we obtain information about the unknown parameters from the observed data of the random variables from the probability distribution governed by the parameters. The Fisher information matrix is a way to quantify the amount of information that the observable random variables carry about the unknown parameters. In our case, the Fisher information matrix is 
\begin{equation} \label{eq:fisher}
\begin{aligned}
  \bm{I}(\bm{\theta}) &  =\mathrm{Var}_{\bm{\theta}} \{\nabla l(\bm{\theta})\}\\
&  = - E_{\bm{\theta}} \{\nabla^2 l(\bm{\theta})\},
\end{aligned}
\end{equation}
  where the notation ``Var'' means the variance, ``E'' means expectation, and   
\begin{equation} \label{eq:h}
  \nabla^2 l(\bm{\theta}) = 
\begin{bmatrix}
N \sum_{k=1}^K x_k^2 g & N \sum_{k=1}^K x_k g\\
N \sum_{k=1}^K x_k g & N \sum_{k=1}^K g
\end{bmatrix},
\end{equation}
  and
\begin{multline} \label{eq:g}
  g=\frac{1}{Nf^2(1-f)^2}\biggl\{-N_{1,k}ff^{\prime} + (N+N_{1,k})f^2f^{\prime} - Nf^3f^{\prime} - Nff^{\prime 2}\\
  +Nf^2f^{\prime 2}  + N_{1,k}ff^{\prime\prime} - (N+N_{1,k})f^2f^{\prime\prime} + Nf^3f^{\prime\prime}\biggr\} .
\end{multline}
  For simpler notation, here $f$ is $f(\alpha x_k + \beta)$; $f^{\prime}$ is $f^{\prime}(\alpha x_k + \beta)$; $f^{\prime\prime}$ is $f^{\prime\prime}(\alpha x_k + \beta)$ and $f^{\prime \prime}(s) = \frac{ d^2 f(s)}{d^2 s}  $

  With the Fisher information matrix, we can also derive the confidence intervals for the MLE\cite{fisherIn}:

\begin{equation} \label{eq:alphaCI}
  \hat{\alpha} \pm z \sqrt{ ( \bm{I}(\hat{\bm{\theta}})^{-1}  )_{11} },
\end{equation}
  and
\begin{equation} \label{eq:betaCI}
  \hat{\beta} \pm z \sqrt{ ( \bm{I}(\hat{\bm{\theta}})^{-1}  )_{22} },
\end{equation}
  where $z$ is the appropriate  critical value (for example, 1.96 for 95 $\%$ confidence), and the notation $( \bm{I}(\hat{\bm{\theta}})^{-1}  )_{ii}$ means that we invert the Fisher information matrix first, and then take the $ii$-th component of the inverted matrix.

\subsection{Signal detection} \label{sec:detection}
The  detection of an exoplanet signal can be modeled as a composite hypotheses testing problem. The null and alternative hypotheses are as follows:
\begin{equation}
\begin{aligned}
& H_0 : \alpha = 0, \beta \geq 0 \\
\text{versus} \quad & \\
& H_1: \alpha > 0, \beta \geq 0 .\\
\end{aligned}
\label{eq:hypothesis}
\end{equation}
Here, since we have no prior information about the exoplanet intensity $\alpha$, and the background light $\beta$,   the posterior probability of the observation under the two hypotheses cannot be calculated to decide which hypothesis is more likely. One solution for this challenge is to use the maximum likelihood. We can conduct a generalized likelihood ratio test (or also called maximum-likelihood test), i.e., we determine whether there is an exoplanet signal based on the ratio
\begin{equation} \label{eq:ratio}
\begin{aligned}
R &= \frac{max_{\alpha, \beta} \ p_1(\bm{y}| \alpha, \beta)}{max_{\alpha, \beta} \ p_0(\bm{y}| \alpha, \beta)}\\
&= \frac{max_{\alpha, \beta} \ \prod_{k=1}^K f(\alpha x_k + \beta)^{N_{1, k}} [1 - f(\alpha x_k + \beta)]^{N_{0, k}}}{max_{\beta} \ \prod_{k=1}^K f(\beta)^{N_{1, k}} [1 - f(\beta)]^{N_{0, k}}}\\
&  = \frac{ \prod_{k=1}^K f( \hat{\alpha_1} x_k + \hat{\beta_1})^{N_{1, k}} [1 - f(\hat{\alpha_1} x_k + \hat{\beta_1})]^{N_{0, k}}}{ \prod_{k=1}^K f(\hat{\beta_0})^{N_{1, k}} [1 - f(\hat{\beta_0})]^{N_{0, k}}}   ,
\end{aligned}
\end{equation}
where $p_0(\cdot)$ and $p_1(\cdot)$ are the probability under hypotheses $H_0$ and $H_1$, respectively, and the two maximal probabilities and the corresponding parameters ($\alpha$ and $\beta$) are calculated using the estimation algorithm in Sec.~\ref{sec:estimation}.  R is the generalized likelihood ratio. For easier calculation, we usually take the log of R, that is the log likelihood ratio. In a real space mission, we can conduct detection while sequentially collecting images. After receiving every new image, we update the test ratio in Eq.~(\ref{eq:ratio}); when $R \geq \pi_U$ or $R \leq \pi_L$, we conclude $H_1$ or $H_0$ is accepted and stop taking new images, otherwise we take a new image and repeat the detection procedure, where $\pi_U$ and $\pi_L$ are thresholds chosen beforehand.

The thresholds   can be determined according to    users' desired true positive   and false alarm rates.   According to Wilks' theorem, the probability distribution of $2R$ under $H_0$, i.e., twice the ratio, is approximately a chi-squared distribution with one degree of freedom\cite{wilks}. There is no simple closed-form theoretical solution for the true positive rate and false alarm rate given a threshold for this model, so we numerically calculate them via Monte Carlo simulation. Given a planet intensity   users want to detect   and the number of PC images that they plan to take,    they can numerically compute the true positive and false alarm rate for each threshold   using this detection algorithm.   They can choose a desirable true positive and false alarm rate value pair and thus its corresponding threshold.

\begin{figure}[ht!]
\centering
\includegraphics[width=\textwidth]{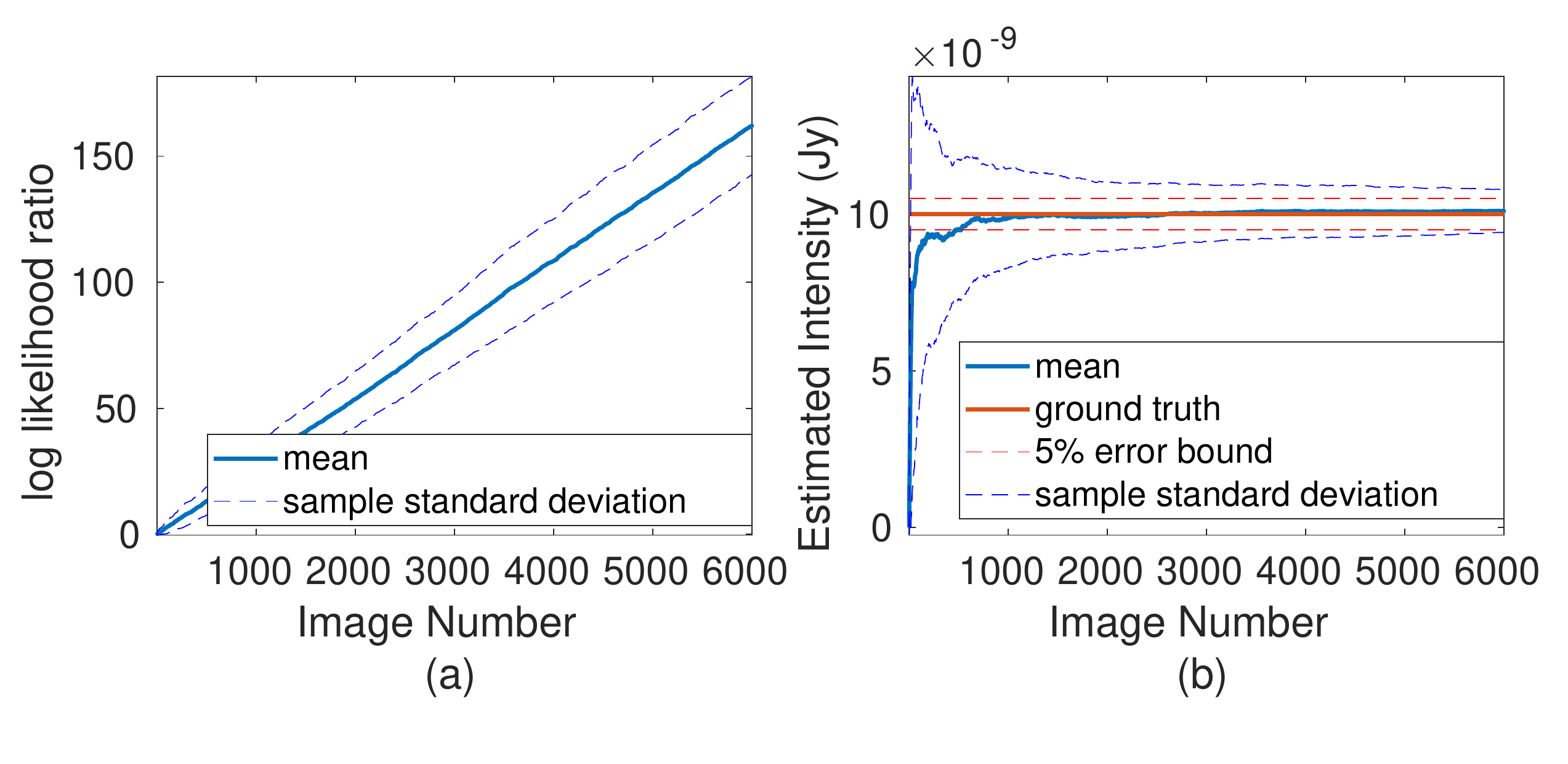}
\caption{ Statistics of the maximum likelihood planet intensity estimation from 100 trials. The   blue solid line is the average of all trials with an increasing number of observations and  the blue dashed lines are the sample standard deviation band. (a) The log-likelihood ratio with an increasing number of observations. The more images that are available, the bigger the log-likelihood ratio is. Thus, the greater the confidence of the existence of a true signal. (b) The estimated intensity with increasing number of images.  The red solid line is the true intensity. The red dashed lines are the 5\% error band. The blue dashed line is the standard deviation band. The method results in a good estimate. }
\label{fig:test}
\end{figure}

     We simulate a signal (signal only) with intensity 1e-8 Jy  (for reference, Venus is 2.99e-8 Jy and Earth is 4.85e-9 Jy if viewed from 10 pc at 0.552~{\textmu}m) and run multiple trials to get the statistics of the method's performance, shown in Fig.\ \ref{fig:test}. The   Bernoulli GLRT method correctly estimates the intensity of the signal quickly. The value of the log-likelihood ratio is high and increases quickly with the increasing number of observations. This means that it is possible to confidently detect the existence of a signal with just a few images and the more observations the larger the gain in confidence for the detection. 

\subsection{Multi-signal detection in an image}\label{sec:whole}
The hypothesis testing process described in the previous section assumes the planet PSF is in the center of the set of $K$ test pixels    (the size of PSF's central core), which will be called an image window from here on. In reality, an image much bigger than the size of a    PSF core will be examined for an unknown number of planets in an unspecified location.  Thus, the detection procedure is repeated for each set of K pixels across the whole image. This generates a log-likelihood ratio map for the image. 

  In this section we demonstrate the performance of the   Bernoulli GLRT detection method on simulated starshade images. The imaging system in all simulations in this paper is the Nancy Grace Roman Space Telescope with a starshade. The pixel size of the detector is 0.021 arcsec by 0.021 arcsec. The optical model to calculate the starshade diffraction uses Fresnel diffraction theory\cite{fresnel} to propagate light past the starshade.  A detailed description of the simulation process can be found in Ref.~\citenum{Mia1}.

\begin{figure}[ht!]
\centering
\includegraphics[width=0.99 \textwidth]{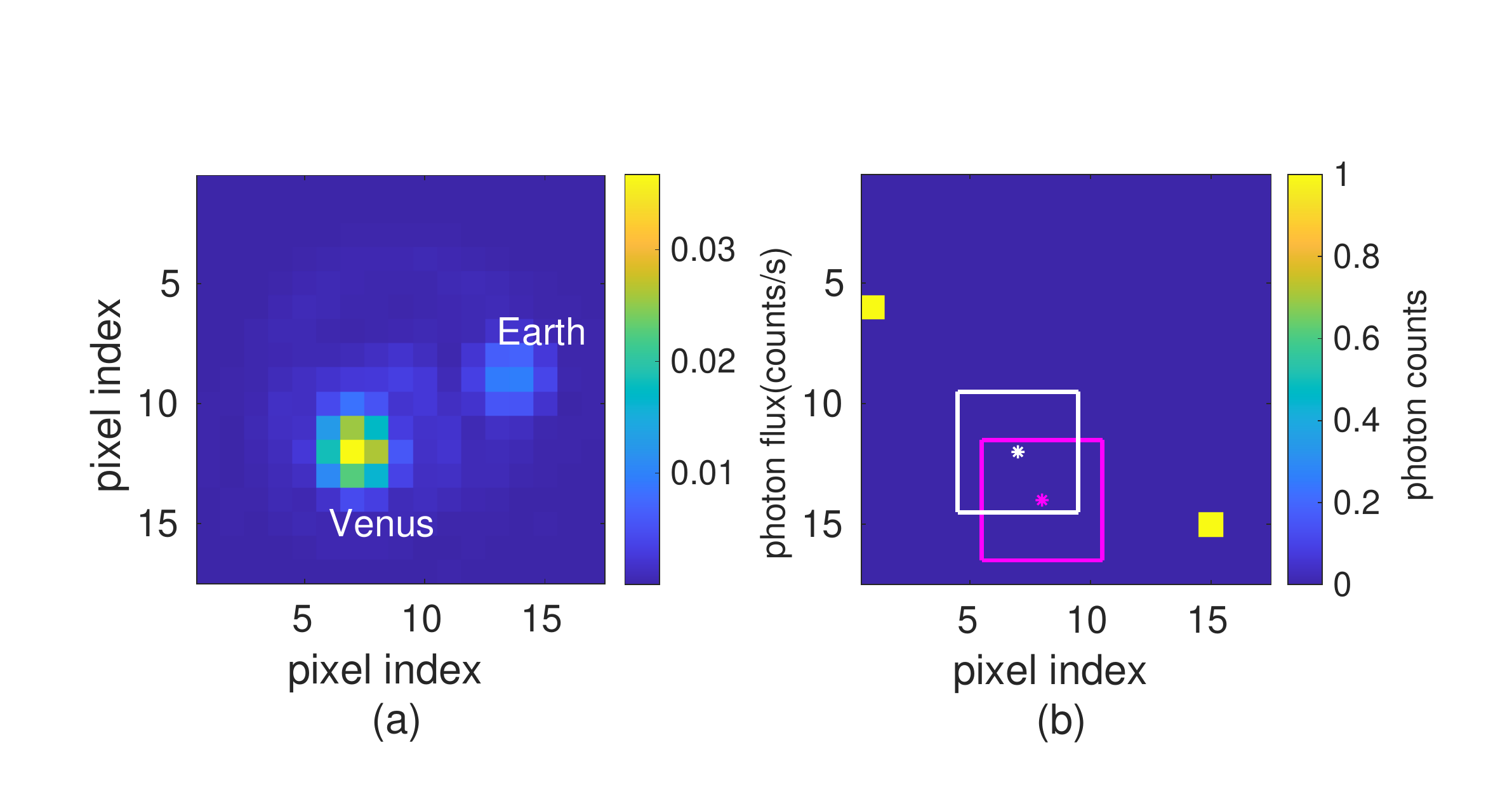}
\caption{ (a) Simulated Noiseless image for the solar system from 10 pc away with a Starshade-Roman Telescope system. This is the ground-truth for the detection problem. Only the Sun, Earth and Venus are included. The brighter signal is Venus and the dim one is Earth. The Sun's light is successfully suppressed by the starshade and thus can hardly be recognized.  (b) One example of a PC image with the detector model described in Section \ref{sec:EMCCD}.   Examples of an image window are also shown. The white box is of the size of PSF core. It is the area used to detect whether there is a planet at the position of the white asterisk, i.e. pixel (7,12). The magenta box is for the magenta asterisk, i.e. pixel (8,14).}
\label{fig:gt}
\end{figure}

  We apply the   Bernoulli GLRT method to simulated starshade images to detect possible exoplanet signals. As shown in Fig.~\ref{fig:gt} (a), we observe our solar system from 10 pc away with a Starshade-Roman Telescope system, where two observable planets are in view, Venus and  Earth. The Sun is present but is not recognizable because the starshade successfully suppressed its light. For this demonstration, we assume no local or extrasolar zodiacal dust. Here we use a fixed integration time (1 s) for each single exposure and wavelength at $\lambda=0.552$~{\textmu}m with a 0.128~{\textmu}m bandwidth. More details about the simulation are available in Ref.~\citenum{Mia1}. After including the photon counting mode detector described in Sec.~\ref{sec:EMCCD},
PC images are produced, an example of which is given in Fig.~\ref{fig:gt} (b).  As the image size is much bigger than the size of a PSF core, the detection procedure is repeated for each image window with the size of a PSF core across the whole image. Examples of an image window are also shown in Fig.~\ref{fig:gt} (b). The white box is of the size of PSF. It is the area used to calculate the likelihood ratio and then decide whether there is a planet at the position of the white asterisk, i.e., pixel (7,12). The magenta box is for the magenta asterisk, i.e., pixel (8,14). Repeating the process, we will get the likelihood ratio at each pixel and thus generates a log-likelihood ratio map for the whole image.

\begin{figure}[ht!]
\centering
\includegraphics[width= \textwidth,trim= 0  0.6cm  0 1.3cm ,clip]{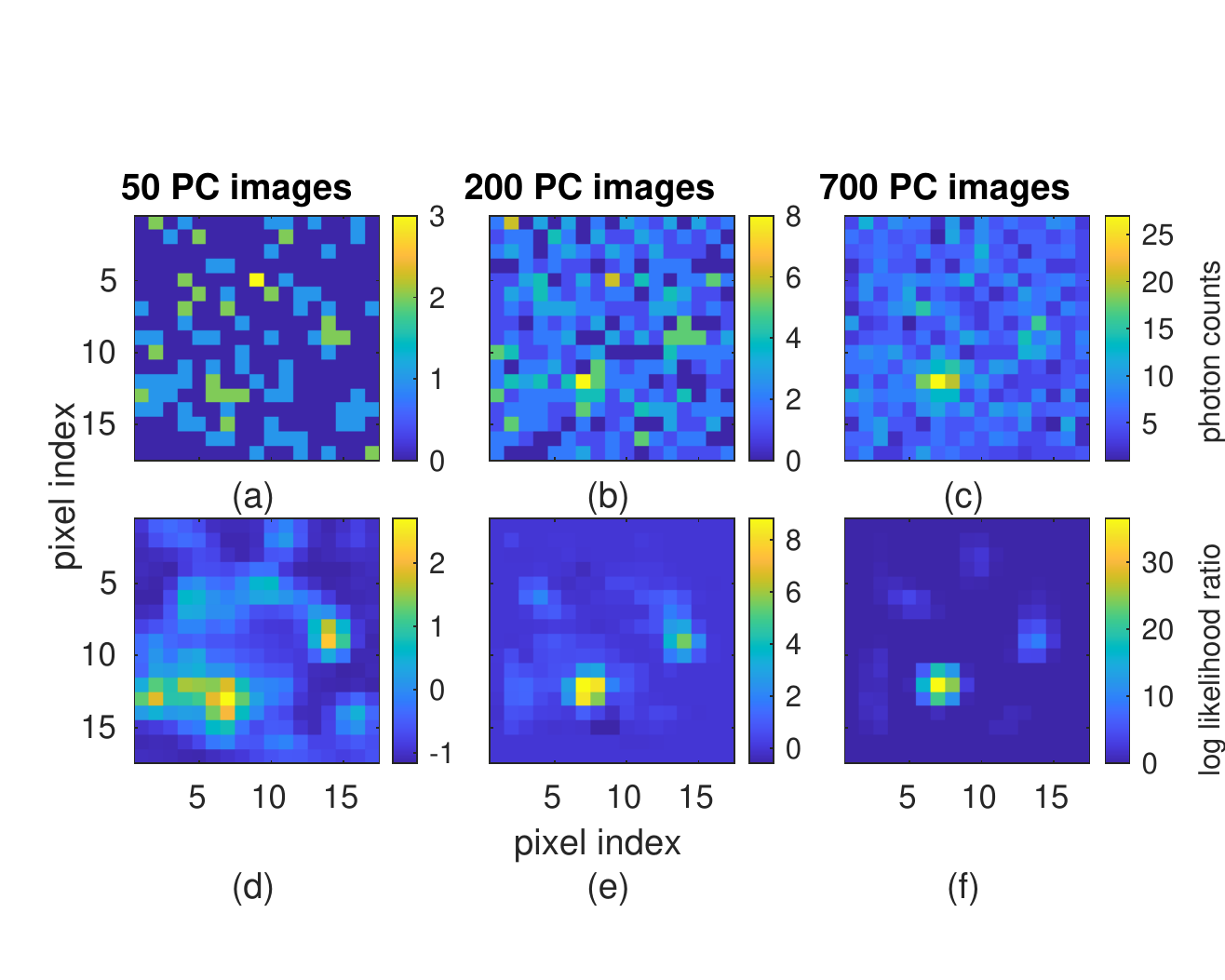}
\caption{Results of the new Bernoulli based GLRT. (a) Co-added image with 50 sequential PC images for Fig.\ref{fig:gt}(a). (b) Co-added image with 200 sequential PC images for Fig.\ref{fig:gt}(a). (c) Co-added image with 700 sequential PC images for Fig.\ref{fig:gt}(a). (d) Log likelihood ratio of each pixel using the 50 sequential PC images in Fig.\ref{fig:nodustimages}(a).  (e) Log likelihood ratio of each pixel  using the 200 sequential PC images in Fig.\ref{fig:nodustimages}(c).  (f) Log likelihood ratio of each pixel  using the 700 sequential PC images in Fig.\ref{fig:nodustimages}(e). The change of Log likelihood ratio of Venus, Earth and a background pixel with increasing number of observations is shown in Fig \ref{fig:llrall}. }
\label{fig:nodustimages}
\end{figure}

\begin{figure}[ht!]
\begin{center}
\begin{tabular}{c}
\includegraphics[width=0.55\textwidth]{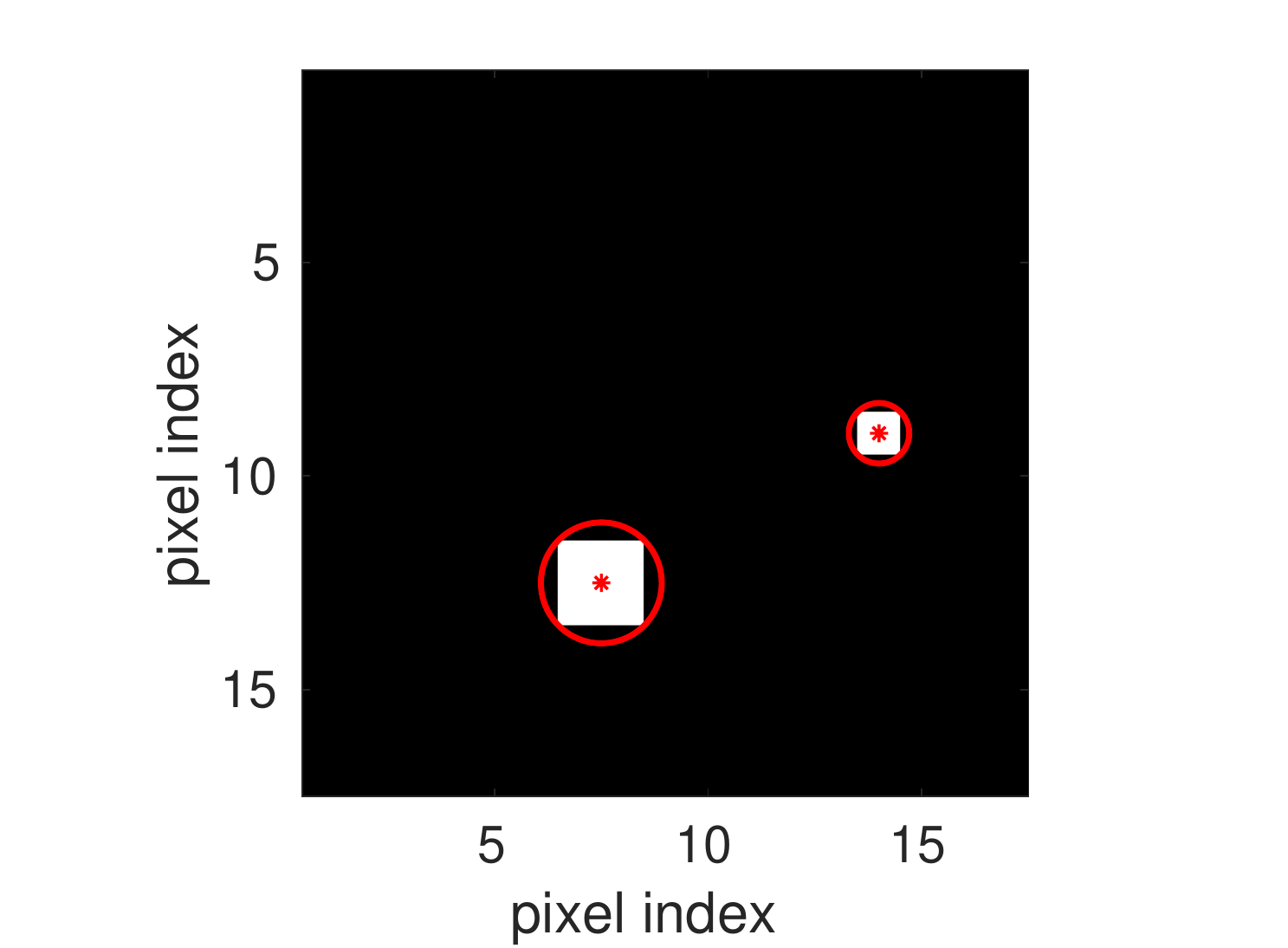}
\end{tabular}
\end{center}
\caption{  Example of position estimation. Binary detection image after thresholding the log-likelihood ratio map in Fig.~\ref{fig:nodustimages}(e) with a threshold of 5 . The red circles are the minimal bounding circles of the detected area. The centers of the circles can be used to estimate the planet positions. \label{fig:pos}}
\end{figure}

  Though the method is applied to each PC image separately rather than on a co-added image, there is insufficient space  to show all PC images one-by-one here.  Thus, to give  a sense of what the data look like, the co-added images are shown in Figs.~\ref{fig:nodustimages}  (a), (b) and (c). Their corresponding log-likelihood ratio maps  given by   Bernoulli GLRT are shown in  Figs.~\ref{fig:nodustimages}  (d), (e) and (f). Venus is located at pixel (7,12), and Earth at (14,9). However, as we see in  Figs.~\ref{fig:nodustimages}  (d), (e) and (f), not only the pixel (7,12), and the pixel (14,9), where the planets are located, have  high log-likelihood ratio values, but the pixels around them also have high log-likelihood ratio values compared to the background. This makes sense as an image window centered on these pixels close to the signal position includes part of the signal. For example, for pixel (8,14), the magenta asterisk, its image window, the magenta box in Fig.~\ref{fig:gt} (b), covers most part of Venus. Our methods only consider two cases. The first is that a signal is at the center of the window. The second one is that no signal exists in the window (only constant background). Therefore, the bright partial signal will have a high log-likelihood ratio. However, the image window only includes part of the signal, so the light distribution doesn't fit  the PSF template perfectly. That is to say, it does not follow Hypothesis 1, so the log-likelihood ratio is lower than that in the real signal pixel. To decide the exact position of the signal, after thresholding the log-likelihood map, we will just choose the center of the detected area's circumscribed circle as the position of the planet. An example is shown in Fig.~\ref{fig:pos}. This planet position estimation should be more robust, especially when only a partial signal is in the image. Our method  detects the existence of the signals quickly  and separates them from the background successfully.  The more observations, the larger the gain in confidence for the detection.

\begin{figure}[ht!]
\centering
\includegraphics[width=0.65 \textwidth,trim= 0  0cm  0 0cm ,clip]{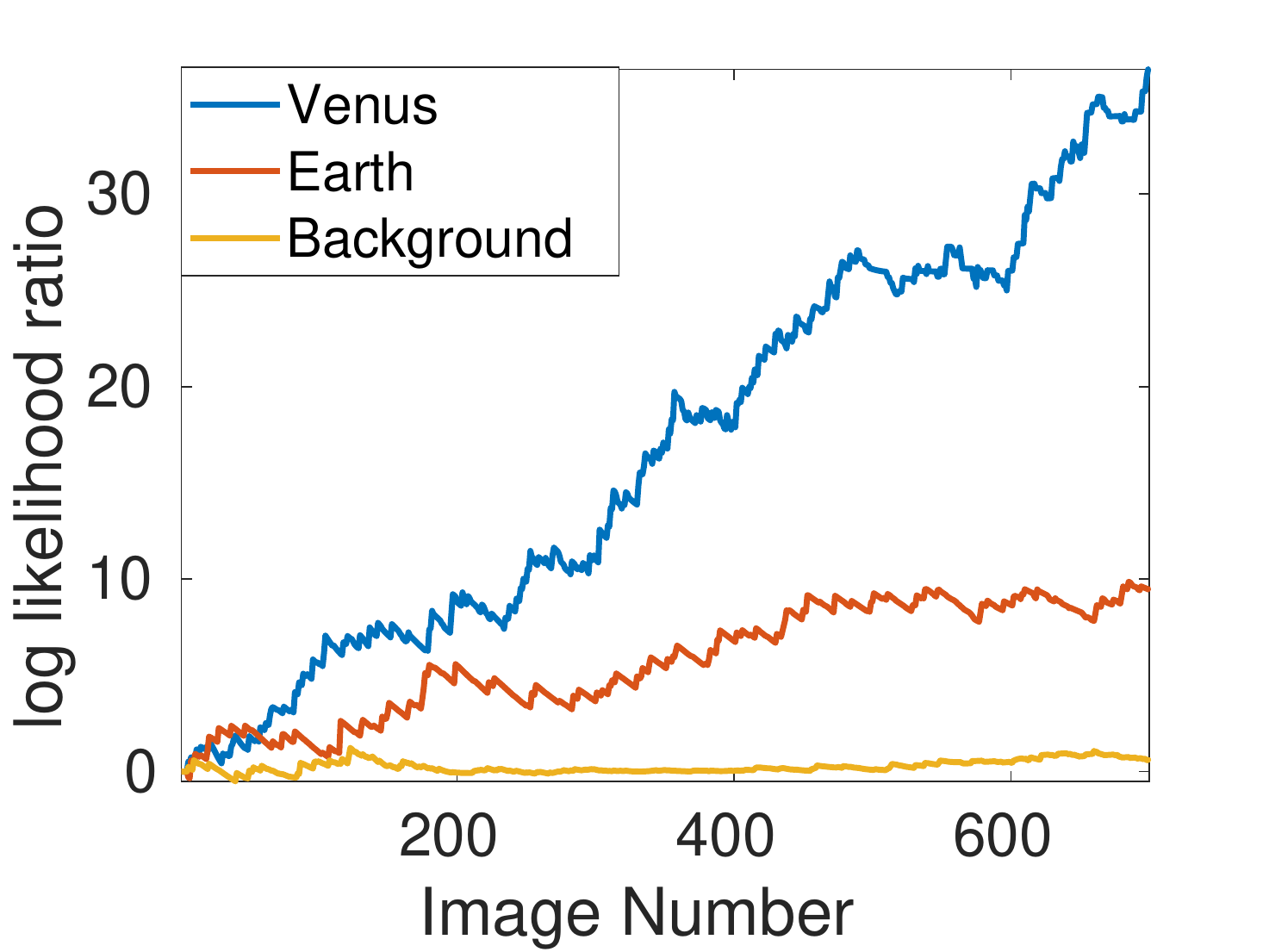}
\caption{ Log likelihood ratio of Venus, Earth and a background pixel at (5,5) with increasing number of observations for Fig.\ref{fig:gt}(a) using Bernoulli based GLRT. }
\label{fig:llrall}
\end{figure}

  The changes of log-likelihood ratio at Venus, Earth and a background pixel at (5,5) after each observation are shown in Fig.~\ref{fig:llrall}. As the figures demonstrate, our method can detect the existence of the signals quickly and separate them from the background successfully. The corresponding false alarm rate for the log-likelihood value of Venus and Earth for the three cases are in Table~\ref{tab:GGLRT}. Details on finding the false alarm rate are given in Sec.~\ref{sec:comp}. The false alarm rate is how likely a background pixel will have a log-likelihood equal or bigger than the chosen threshold value and thus get wrongly considered as a detection. The smaller the false alarm rate is, the more confidence we have for the detection. As shown by the example in this section, the more observations, the more confidence we gain for the detection. 

  When two signals overlap, the light distribution in the search area centered on the signals will be distorted. This violates the hypothesis that the image area contains a PSF-shaped signal and constant background. Therefore, the estimation and detection performance may be influenced. If there is not too much  overlap, the intensity estimate for both signals will be higher than the true values and the position estimate will biased towards each other. If the signals are close enough, the algorithm may consider them as one signal. Future work can look into solving this problem by expanding the image area and modeling overlapping signals. Moreover, astronomers can take another set of images after a while, because the signals' relative positions are likely to change and will be separate. In this work, we will not further consider the problem.

   In the case of non-uniform background, the algorithm's performance depends on how well we know the non-uniform background. The current algorithm finds the signal looking like the PSF template against the uniform background in an area of the size as PSF core. We traverse the whole image by checking the PSF-core sized area one by one. Thus, the assumption of the algorithm is that the background is locally constant but can be non-uniform in the whole image. If the non-uniform background changes slowly spatially, a constant background may still be a good approximation and the detection will not be influenced much. If we have zero knowledge about the distribution of the bright background close to the planets, the results are affected by the distorted signal. However, if we have some knowledge about the nonuniform background, we can accommodate it in the model. For example, it is a reasonable approximation to assume the face-on exozodiacal light  being axisymmetric. We have developed an iterative GLRT to do detection and estimation\cite{gglrt}, which is essentially an Expectation-Maximization (EM) algorithm. We iteratively estimate either the planets' signal or the exozodiacal dust first, and then use the estimation as a known factor to estimate the other until both estimates converge\cite{gglrt}.

\subsection{Comparison with other methods}
\label{sec:comp} 
In Ref.~\citenum{Mia,gglrt}, we introduced a GLRT method that was applied to co-added images generated from the combination of many frames. That method assumed the values in each pixel followed a Gaussian distribution, so the summation of frames was necessary in order to build enough signal in each pixel for the Gaussian assumption to be valid. Alternatively, the   Bernoulli distribution in the new BGLRT method works for single PC images, which makes it easier to use online as it can process    even a few PC images. 

\begin{figure}[ht!]
\centering
\includegraphics[width=0.99 \textwidth,trim= 0  1cm  0 0cm ,clip]{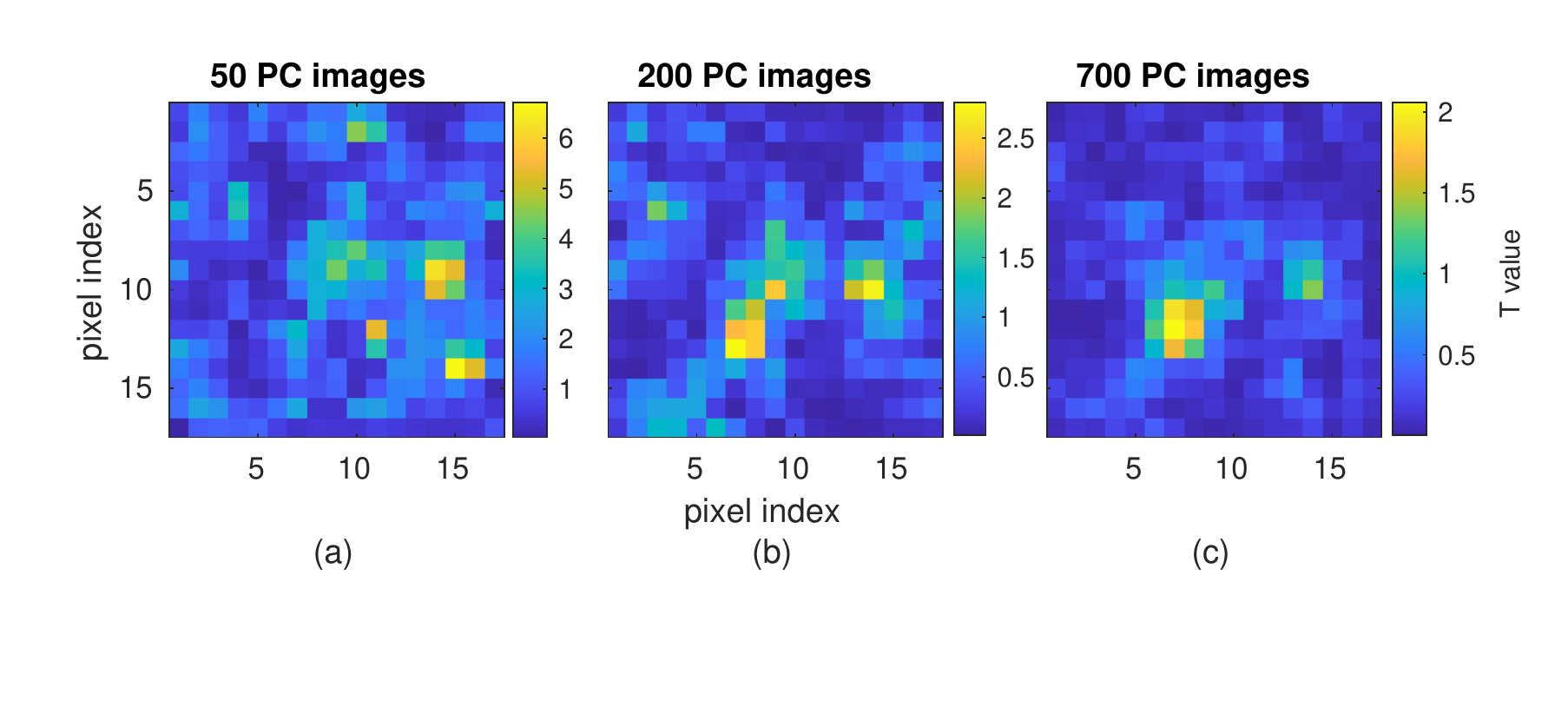}
\caption{Results of a Gaussian based GLRT for images in Fig.\ref{fig:nodustimages}.  (a) T value of each pixel for  the co-added image with 50 sequential PC images for Fig.\ref{fig:nodustimages}(a). (b) T value of each pixel for  the co-added image with 200 sequential PC images for Fig.\ref{fig:nodustimages}(b). (c) T value of each pixel for  the co-added image with 700 sequential PC images for Fig.\ref{fig:nodustimages}(c).  }
\label{fig:example2}
\end{figure}

The underlying relationship between the probability in the Bernoulli distribution and flux intensity can be derived theoretically based on the detector model or directly measured in experiment. No approximation is used and thus the method is accurate and efficient at extracting information. In Figs.~\ref{fig:example2} (a),(b) and (c) we show the T map using the Gaussian GLRT for the co-added images of Fig.~\ref{fig:nodustimages}\cite{Mia}.   The T value is a proxy for the log-likelihood ratio that is easier to compute. The T value in a pixel is given by 
\begin{equation}
\label{T}
 T(\bm{x}_{k})  = (N-2) ( L_G (\bm{x}_{k})^{\frac{2}{N}}-1)\,,
\end{equation}
where  $L_G (\bm{x}_{k})$ is the log-likehood ratio    of the co-added image $\bm{x}_{k}$ under a Gaussian assumption. The T value consists of basic operations (matrix multiplication, summation, division) of the image and parameters, so it is easier to calculate than $L_G$\cite{Mia,gglrt}. The T value is used to compared with a threshold to get a detection result in the Gaussian GLRT. Fig.~\ref{fig:example2} shows that the T map is noisier than the log-likelihood map from the   Bernoulli GLRT. The corresponding false alarm rate for the T value of Venus and Earth for the three cases are in table~\ref{tab:GGLRT}. We generally have higher confidence using   Bernoulli GLRT.

\begin{figure}[ht!]
\centering
\includegraphics[width=0.99 \textwidth,trim= 0  1cm  0 0cm ,clip]{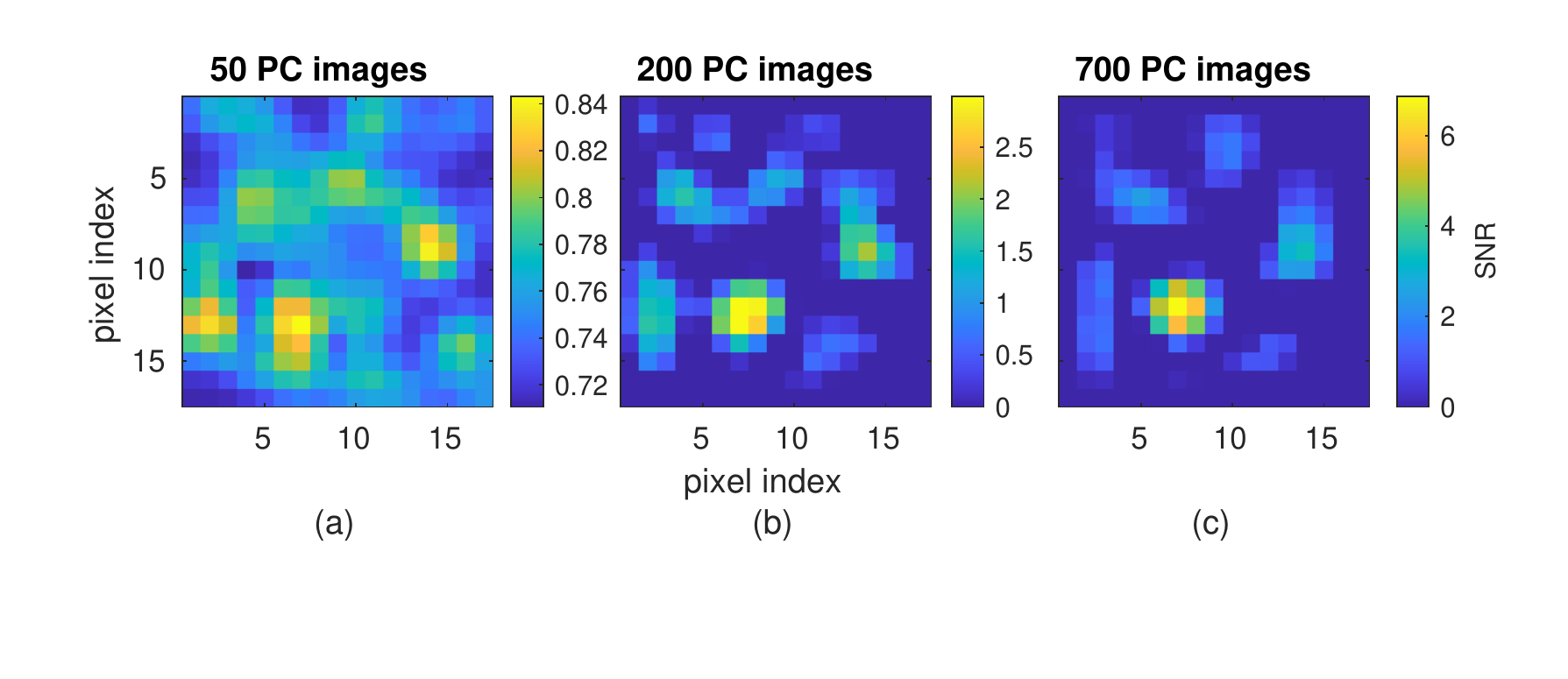}
\caption{ The SNR map based on the  Bernoulli model for images in Fig.~\ref{fig:nodustimages}.  (a) The SNR map of each pixel for  the co-added image with 50 sequential PC images for Fig.~\ref{fig:nodustimages}(a). (b) The SNR map of each pixel for  the co-added image with 200 sequential PC images for Fig.~\ref{fig:nodustimages}(b). (c) The SNR map of each pixel for  the co-added image with 700 sequential PC images for Fig.~\ref{fig:nodustimages}(c).  } 
\label{fig:fisher}
\end{figure}

  The above comparison indicates that the Bernoulli model for PC images contributes to the performance improvement. To further analyze the contribution from GLRT, we also compared the Bernoulli GLRT with the performance of detection based on SNR from our Bernoulli model. With the MLE of signal intensity and estimated standard deviation of the MLE derived from the Fisher information matrix in Sec.~\ref{sec:estimation}, we define their ratio as the SNR from our model. We will refer to this SNR definition as Bernoulli SNR (BSNR). For each pixel, we utilize the image window centered at it and calculate the estimated signal intensity and the standard deviation as in  Sec.~\ref{sec:estimation}. Then, we calculate the ratio. After repeating the process for all the pixels in the image, we obtain a SNR map. The SNR maps for the three example co-added images in Fig.~\ref{fig:nodustimages} are shown in Fig.~\ref{fig:fisher} to help visually compare the performance with the BGLRT method.

\begin{table}[ht]
\caption{ False alarm rate (FA) for the cases in Fig.~\ref{fig:nodustimages} using      Bernoulli GLRT (BGLRT) and Gaussian GLRT (GGLRT) }
\label{tab:GGLRT}
\begin{center}       
\begin{tabular}{|l|l|l|l|l| } 
\hline
\rule[-1ex]{0pt}{3.5ex} Case & \begin{tabular}{@{}c@{}} Venus  \\  BGLRT   \end{tabular}  &  \begin{tabular}{@{}c@{}} Venus  \\ GGLRT  \end{tabular}   &    \begin{tabular}{@{}c@{}} Earth \\ BGLRT  \end{tabular}      &    \begin{tabular}{@{}c@{}} Earth \\ GGLRT \end{tabular}  \\
\hline
\rule[-1ex]{0pt}{3.5ex}  Fig.~\ref{fig:nodustimages}(a) &  0.326 &  0.655& 0.076&  
    0.110 \\
\hline
\rule[-1ex]{0pt}{3.5ex} Fig.~\ref{fig:nodustimages}(b)& 0.000 &   0.002 &  2.000e-04 &    0.004  \\
\hline
\rule[-1ex]{0pt}{3.5ex} Fig.~\ref{fig:nodustimages}(c)& 0.000 &0.000  &  0.000&	  0.000 \\
\hline
\end{tabular}
\end{center}
\end{table}

\begin{figure}[ht!]
\centering
\includegraphics[width=0.99 \textwidth,trim= 0  4cm  0 1cm ,clip]{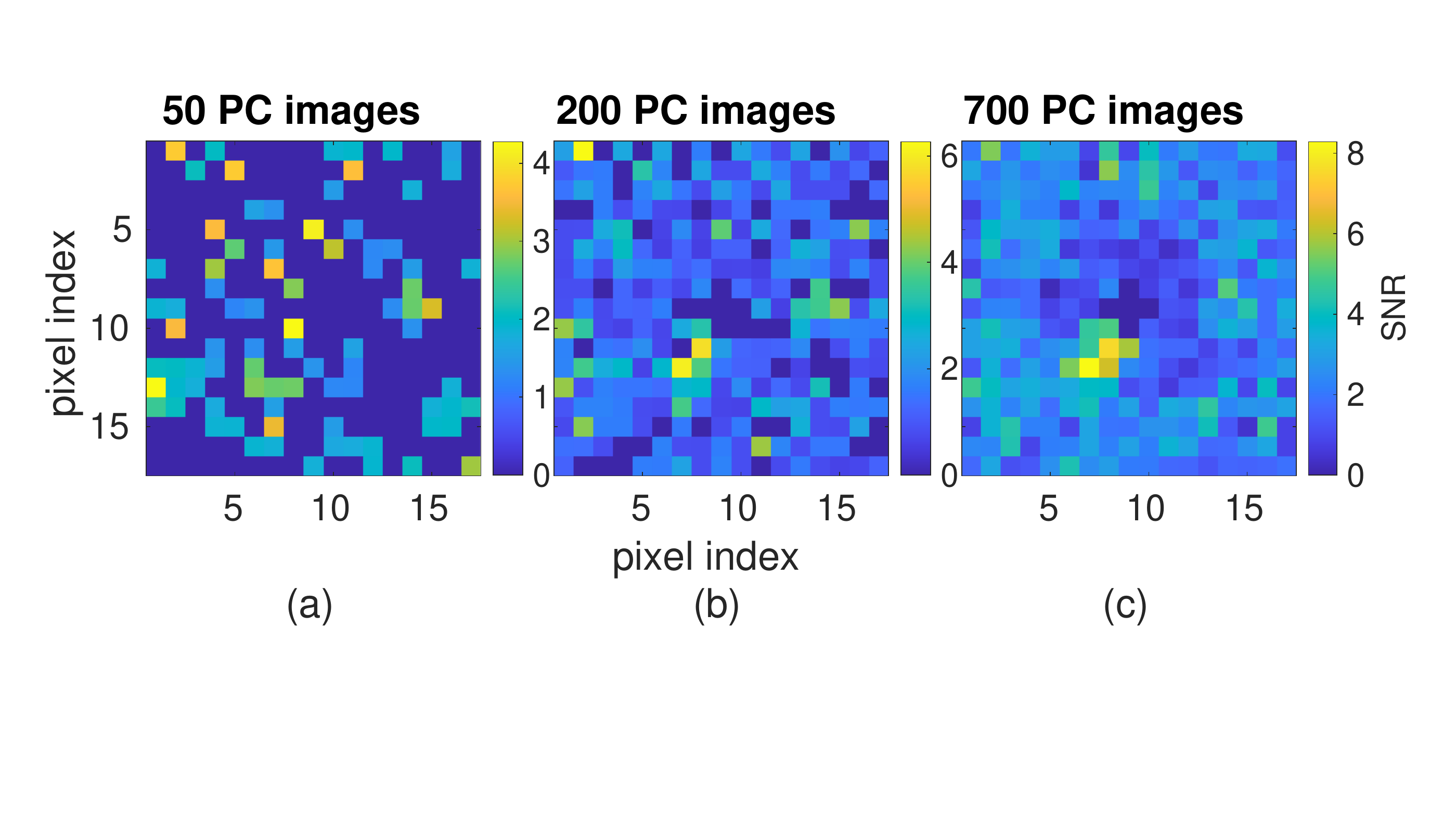}
\caption{The SNR map from the   pyKLIP package\cite{pyklip} for images in Fig.\ref{fig:nodustimages}. (a)  SNR of each pixel for  the co-added image with 50 sequential PC images   as shown in Fig.\ref{fig:nodustimages}(a). (b) SNR  of each pixel for  the co-added image with 200 sequential PC images   as shown in Fig.\ref{fig:nodustimages}(b). (c) SNR of each pixel for  the co-added image with 700 sequential PC images   as shown in Fig.\ref{fig:nodustimages}(c).  }
\label{fig:VIP}
\end{figure}

We also compared the   Bernoulli GLRT with the performance   of the detection method based on   the SNR map implemented in pyKLIP\cite{pyklip}.   For each pixel, the algorithm masks   its surrounding pixels within the signal area in question (We chose the size of signal area as the size of PSF core) and then computes the standard deviation using the rest of pixels in concentric annuli.    The ratio of the pixel value and this standard deviation as SNR value at this pixel. The width of the annuli used is the diameter of the PSF core in this paper.   Repeating the process for all the pixels in the image generates a SNR map. The SNR maps for the three example co-added images in Fig.~\ref{fig:nodustimages} are shown in Fig.~\ref{fig:VIP} to help visually compare the performance with the GLRT methods.

\begin{figure}
\centering
  \includegraphics[width=\linewidth,trim= 0  4.5cm  0 2.5cm ,clip]{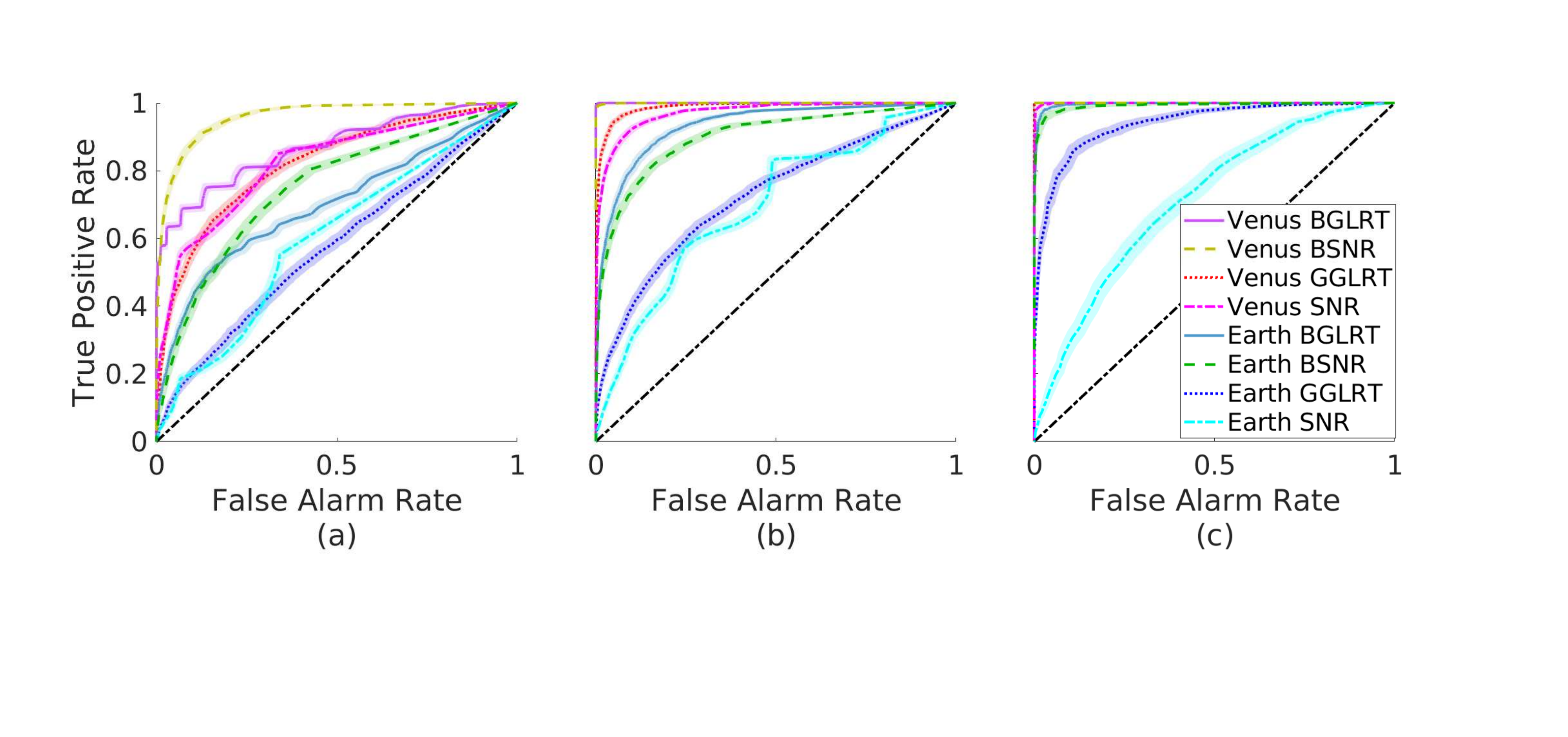}
\caption{Receiver operating characteristic curves  with confidence interval for Earth and Venus detection with the   four different methods. `BGLRT' means the result used the   Bernoulli GLRT described in this paper, which processes    each PC images   sequentially using   the GLRT based on a Bernoulli distribution. `BSNR' means the detection method based on our Bernoulli SNR, which utilizes the final MLEs from our Bernoulli model and estimated standard deviation from Fisher information matrix. `GGLRT' in the legends means the result used the previous GLRT\cite{Mia}, which processes co-added images and assumes Gaussian noise. `SNR' means the detection method based on SNR implemented in pyklip   , which is applied to co-added images. The shaded region behind each ROC curve is its 95$\%$ confidence interval. (a) ROC curves using 50 PC images calculated from 5000 trials. (b) ROC curves using 200 PC images calculated from 5000 trials. (c) ROC curves using 700 PC images calculated from 2000 trials.}
\label{fig:roc}
\end{figure}

To further demonstrate the different properties of the methods, we compare their Receiver Operating Characteristic (ROC) curves for the detection of Venus and Earth. It is difficult to analytically derive a closed form relationship between the false alarm rate and true positive rate and the threshold chosen in the new Binomial GLRT, unlike our previous GLRT\cite{Mia}, which assumes Gaussian noise. Thus, we use Monte Carlo simulations, using multiple thresholds for each simulation, to calculate the ROC curves, shown in Fig.~\ref{fig:roc}.   The simulation is the same as described in Sec.~\ref{sec:whole}, which is the starshade images with the Sun, Venus and Earth.  We compare the performance of the methods using different numbers of total images, which is denoted as $n_{pc}$. We apply the   Bernoulli GLRT on the set of images and obtain the log-likelihood ratio map; we also apply the Gaussian GLRT to get the false alarm rate map.   We also apply Bernoulli SNR and SNR from pyKLIP on the images. We apply a set of different thresholds with the resulting detection or missed detection of Earth, Venus and a background pixel. We run a large number of trials, which is denoted as $n_{trials}$, and record the ratio of detection of Earth and Venus as the true positive rate for Earth and Venus, and record the ratio of detection of the background pixel as the false positive rate.   For each ROC curve, we also calculate the confidence interval shown as the shaded areas in Fig.~\ref{fig:roc}. The confidence interval is standard Monte Carlo statistics, which is detailed in the Appendix. As can be seen in the Fig.~\ref{fig:roc}, the confidence interval is tight.

For Fig.~\ref{fig:roc}(a), we ran 5000 trials with  50 PC images generated for Venus and Earth in each trial. The   Bernoulli GLRT and   Bernoulli SNR were applied to the  50 PC images   sequentially; the Gaussian GLRT and   SNR from pyKLIP were  applied to the co-added image of the 50 PC images. Results for different decision thresholds were recorded and combined into the ROC curve. For Fig.~\ref{fig:roc}(b), we ran 5000 trials and 200 PC images for each trial. For Fig.~\ref{fig:roc}(c), we ran 2000 trials and 700 PC images for each trial. The performance for Venus is better than Earth when we use the same method and the same number of PC images. One reason is that Venus is brighter than Earth. Moreover, Venus is further away from the star at the center, so the influence from the residual starlight is smaller compared to that for Earth. The performance for both Venus and Earth is better with more images. It is easy to understand that the method performs better with more information, i.e. more images.    Generally speaking, the   Bernoulli GLRT outperforms the other methods.

  For easier comparison, we also list the area under the ROC curve (AUC) for all the curves in Table.~\ref{tab:auc}.  AUC is an aggregate measure of performance across all possible thresholds. It can be interpreted as the probability that the model ranks a random positive example higher than a random negative example. AUC is 1 if the model's decisions are all correct and 0 if all wrong.  More comparisons of GLRT with Gaussian assumption and the SNR method can be found in our other work\cite{gglrt}. Overall,   Bernoulli GLRT outperforms GLRT with Gaussian assuption and the SNR method.

\begin{table}[ht]
\caption{   Comparison of area under the curve (AUC) for   BGLRT, BSNR, GGLRT and SNR method from pyKLIP\cite{pyklip}. }
\label{tab:auc}
\begin{center}       
\begin{tabular}{|l|l|l|l|l|l|l|l|l|} 
\hline
\rule[-1ex]{0pt}{3.5ex}   &  \begin{tabular}{@{}c@{}} Venus \\ BGLRT \end{tabular} &  \begin{tabular}{@{}c@{}}   Venus \\   BSNR \end{tabular} &\begin{tabular}{@{}c@{}} Venus \\GGLRT\end{tabular}&\begin{tabular}{@{}c@{}} Venus \\SNR\end{tabular} &  \begin{tabular}{@{}c@{}}Earth\\BGLRT \end{tabular}  &  \begin{tabular}{@{}c@{}}   Earth\\  BSNR \end{tabular}   &\begin{tabular}{@{}c@{}}Earth\\GGLRT\end{tabular} &\begin{tabular}{@{}c@{}}Earth\\SNR\end{tabular}   \\
\hline

\rule[-1ex]{0pt}{3.5ex} 50PC &0.8697&   0.9594 &0.8205 & 0.8224&0.6991&  0.7488 & 0.5753&0.5991 \\
\hline
\rule[-1ex]{0pt}{3.5ex} 200PC &	0.9999&  0.9996  &  0.9878& 0.9714 & 0.9392 &   0.9002 & 0.7245 &0.6953\\
\hline
\rule[-1ex]{0pt}{3.5ex} 700PC &	1 &  1   & 1&  0.9993   &0.9970 &  0.9920& 0.9401 &   0.7159\\
\hline
\end{tabular}
\end{center}
\end{table}

\subsection{Early stopping for observation when no planets exists}

As shown in the previous examples, the   Bernoulli GLRT successfully detects a planet even for short integration times. Another important problem is to know when to stop when no detection is made so time wasted on a system with no planet is minimized. As no closed-form analytical solution for   Bernoulli GLRT can be derived among false alarm rate and intensity and number of PC images, which will define the total time needed, it is impossible to solve for the time needed to reach a specific false alarm rate given the planet intensity via inverting a function. Instead, we will rely on numerical calculation via Monte Carlo simulation.

\begin{figure}
\centering
  \includegraphics[width=0.7\linewidth]{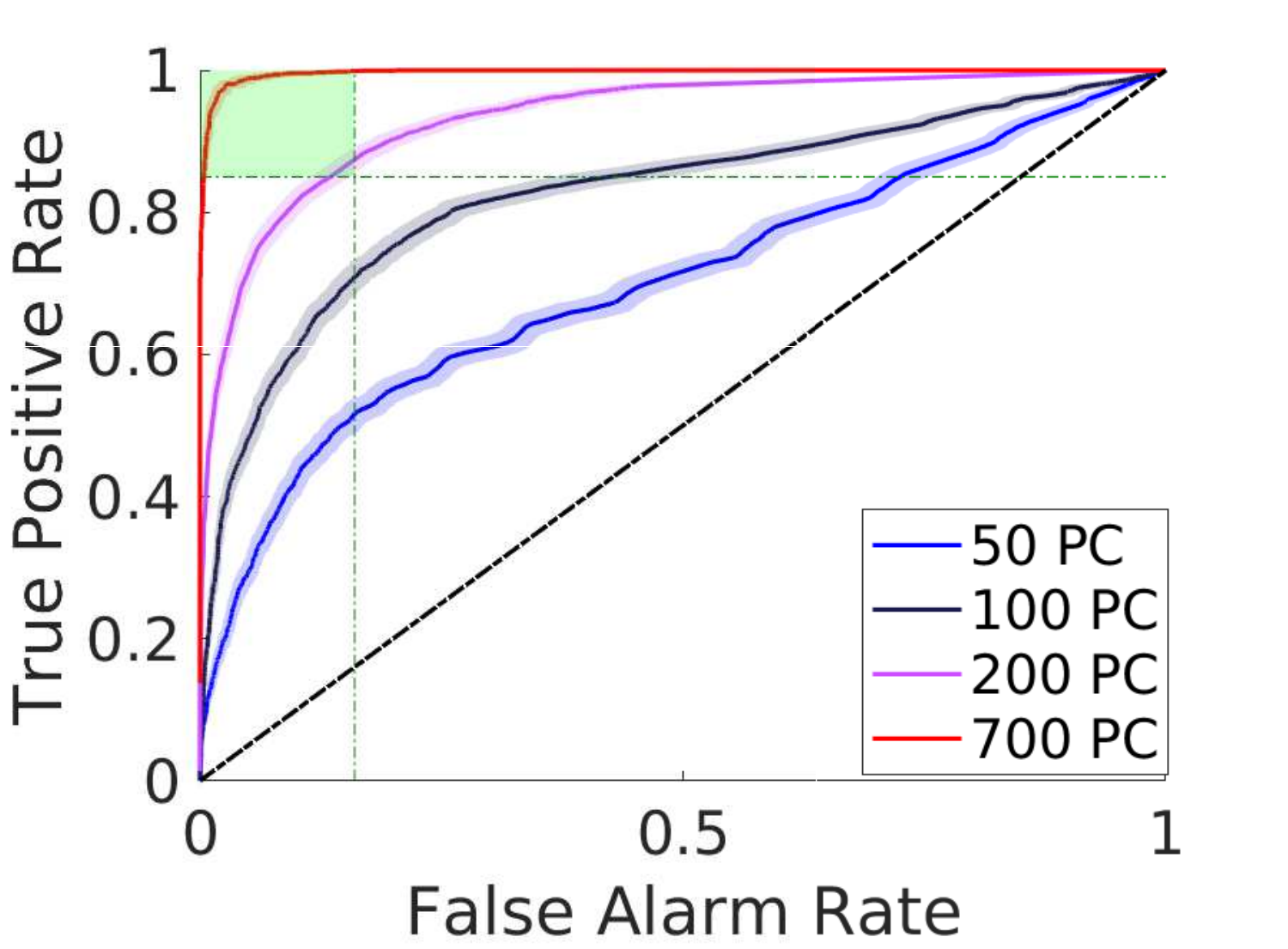}
\caption{ ROC with confidence interval for Earth using   Bernoulli GLRT with different number of PC images. The horizontal green dotted line is the acceptable true positive rate (TP) and the vertical one is the acceptable false alarm rate (FA) region. Thus, the acceptable true TP   and acceptable  FA region is the shaded green area. The ROC with 200 PC is the first ROC reaching the acceptable region with the fewest possible images.}
\label{fig:stop}
\end{figure}

First, three parameters are specified: the minimum planet intensity to be detected, the maximum false alarm rate that can be accepted, and the minimum true positive rate that is acceptable. Then, for a different number of images, the ROC is calculated via monte carlo simulation. Finally, the minimum number of PC images that can reach the requirements are chosen. For example, if we assume the dimmest planet has the same intensity as Earth, the maximum acceptable false alarm rate is 0.16 and the minimum acceptable true positive rate is 0.85. The acceptable false alarm rate, true positive rate pairs are in the shaded green region in Fig.~\ref{fig:stop}. We calculate the ROCs with different numbers of PC images and find that 200 PC images' ROC is the first one to reach the green region, as shown in  Fig.~\ref{fig:stop}. Thus, after taking 200 PC images and if there is still no signal detected, the telescope system could move on to observing another target star with confidence there is no planet.

\section{Conclusions}
In this paper we presented a model and signal processing approach based directly on single photon counting images. Previous work on PC image processing for the most part only uses co-added PC images rather than operating on individual PC images. Such methods do not provide theoretical guidance for choosing the total number of PC images to combine into one co-added image   as in our previous work on GLRT under Gaussian assumption\cite{gglrt}. Our method, the BGLRT, here directly works with each PC image, so there is no need to choose the hyper parameter, the number of PC images to combine into one co-added image. Furthermore, previous work on PC image processing assumes Gaussian noise for co-added images, but the method here directly uses the detector model, which more accurately represents its noise characteristics. In this paper, we showed that such a method outperforms our previous GLRT method using co-added images under a Gaussian noise assumption, and   the SNR method commonly used in high-contrast imaging.    We also compare the BGLRT with the BSNR, which also uses the estimation results from our Bernoulli model. The BGLRT and the BSNR have similar performance (The BGLRT is a little better in most cases) and are better than other methods, which indicates the performance gain in detection is mostly the result of improved model for the imaging process.    Furthermore, our method provides the maximum likelihood estimate of exoplanet intensity and background intensity.

This approach maximizes the utilization of information presented in each observation. We applied this method to simulated starshade images and successfully detected Venus and Earth. Directly processing the PC images online helps allocate the observation time efficiently.  We can compare the log-likelihood ratio with thresholds chosen beforehand after each observation and stop accordingly. As we can decide the existence or lack of planet efficiently in this way, we can move to other planetary systems if there is no planet or make sure that enough information is gathered if there is a planet.   Besides the observation time, the analysis of detection performance introduced could also give quantitative guidance on the choice of imaging parameters, such as the threshold.
 
   The Bernoulli model for PC images and the resulting GLRT for hypothesis testing are a general model and detection method that can be significantly expanded for different applications. This work focuses on the simple case to demonstrate the concept, but the framework is more general. Through Eq.~\ref{eq:probI}, the likelihoods in different pixels are related by the underlying template for signals we want to detect, i.e. a PSF template plus constant background as in Eq.~\ref{eq:signal}. Thus, the resulting  GLRT decides whether such template exists in the observed area. However, this is just one particular application example of the model and the method. By replacing $x$ in Eq.~\ref{eq:signal}, different signal templates can be chosen to appropriately match the observation scenario; the remaining steps in the derivation and the application of the   Bernoulli GLRT are the same.

\appendix    

\section{Confidence interval for ROC curves}
\label{appendix}
  The confidence interval of a proportion $\hat{p}$ is given by \cite{statistics}:
\begin{equation}
  \left(\hat{p} - z_{1- \eta/2} \sqrt{ \frac{\hat{p}(1-\hat{p})}{n_{trials}} },~\hat{p} + z_{1-\eta/2} \sqrt{ \frac{\hat{p}(1-\hat{p})}{n_{trials}} } \right)\,. 
\label{eq:ci}
\end{equation}
   This confidence interval is based on Fisher information. Thus, for a point $(\hat{p}_{false}, \hat{p}_{positive})$ on the ROC curve, we take the confidence interval using the two points:
\begin{equation}
  \left(\hat{p}_{false} - z_{1-\eta/2} \sqrt{ \frac{\hat{p}_{false}(1-\hat{p}_{false})}{n_{trials}} },~\hat{p}_{positive} + z_{1-\eta/2} \sqrt{ \frac{\hat{p}_{positive}(1-\hat{p}_{positive})}{n_{trials}} } \right)
\label{eq:cil}
\end{equation}
  and
\begin{equation}
  \left(\hat{p}_{false} + z_{1-\eta/2} \sqrt{ \frac{\hat{p}_{false}(1-\hat{p}_{false})}{n_{trials}} },~\hat{p}_{positive} - z_{1-\eta/2} \sqrt{ \frac{\hat{p}_{positive}(1-\hat{p}_{positive})}{n_{trials}} } \right)\,
\label{eq:cr}
\end{equation}  
  where $z_{1- \eta}/2$ is the upper critical value for the standard normal distribution with $ \eta/2$ area to the right of it.

\subsection*{Disclosures}
The authors have no relevant financial interests in the manuscript and no other potential conflicts of interest to disclose.

\subsection* {Acknowledgments}
This work is supported by Caltech-JPL NASA grant NNN12AA01C. The authors would like to thank the anonymous reviewers for the insightful comments,    especially for suggesting the comparison with methods based on SNR defined with our Bernoulli model results.


\bibliography{report}   
\bibliographystyle{spiejour}   


\vspace{2ex}\noindent\textbf{Mengya Hu} is a Ph.D. candidate in Mechanical and Aerospace Engineering, Princeton University. Her research focuses on the image simulation and signal detection of space telescope systems with starshades. She graduated from Department of Thermal Science and Energy Engineering, University of Science and Technology of China in 2015 and was awarded the highest honor of the university, the “Guo Mo-Ruo Scholarship”.

\vspace{2ex}\noindent\textbf{He Sun} is a postdoctoral researcher in Computing and Mathematical Sciences (CMS) at the California Institute of Technology. He received his PhD from Princeton University in 2019 and his BS degree from Peking University in 2014. His research focuses on adaptive optics and computational imaging, especially their applications in astrophysical and biomedical sciences, such as exoplanet and black hole imaging.

\vspace{2ex}\noindent\textbf{Anthony Harness} is an Associate Research Scholar in the Mechanical and Aerospace Engineering Department at Princeton University. He received his Ph.D. in Astrophysics in 2016 from the University of Colorado Boulder. He currently leads the experiments at Princeton validating starshade optical technologies. 

\vspace{2ex}\noindent\textbf{ N. Jeremy Kasdin} is the Assistant Dean for Engineering at the University of San Francisco and the Eugene Higgins professor of Mechanical and Aerospace Engineering, emeritus, at Princeton University. He received his Ph.D. in 1991 from Stanford University. After being the chief systems engineer for NASA's Gravity Probe B spacecraft, he joined the Princeton faculty in 1999 where he researched high-contrast imaging technology for exoplanet imaging. From 2014 to 2016 he was Vice Dean of the School of Engineering and Applied Science at Princeton. He is currently the Adjutant Scientist for the coronagraph instrument on NASA’s Wide Field Infrared Survey Telescope.

\listoffigures
\listoftables

\end{spacing}
\end{document}